\documentstyle[amssymb,epsfig,12pt]{article}
\newcommand{\be}{\begin{equation}}
\newcommand{\bea}{\begin{eqnarray}}
\newcommand{\ee}{\end{equation}}
\newcommand{\eea}{\end{eqnarray}}
\newcommand{\ba}{\begin{array}}
\newcommand{\ea}{\end{array}}

\def\ds{\displaystyle}
\def\noi{\noindent}
\def\ol{\overline}
\def\pa{\partial}
\def\f{\frac}

\def\ga{\gamma}
\def\a{\alpha}

\def\p{\varphi}
\def\P{\Phi}
\def\d{\delta}
\def\ka{\kappa}
\def\ga{\gamma}
\def\z{\zeta}
\def\b{\beta}

\def\t{\tilde}
\def\L{\Lambda}
\def\l{\lambda}

\def\mb{\mathbf}
\def\na{\nabla}
\begin{document}

\title{\bf Functional Renormalization Description of the Roughening Transition}

\vskip 3 true cm

\author{Anusha Hazareesing$^{1,2}$ and Jean-Philippe Bouchaud$^1$}

\date{\it $^1$ Service de Physique de l'\'Etat Condens\'e,
 Centre d'\'Etudes de Saclay, \\ Orme des Merisiers, 
91191 Gif-sur-Yvette Cedex, France \\ 
$^2$ Laboratoire de Physique Th\'eorique de l'Ecole Normale Sup\'erieure
 \footnote {Unit\'e propre du CNRS,  associ\'ee
 \`a\ l'Ecole
 Normale Sup\'erieure et \`a\ l'Universit\'e de Paris Sud} , \\
24 rue
 Lhomond, 75231 Paris Cedex 05, France }


\maketitle

\begin{abstract}
We reconsider the problem of the static thermal roughening of an elastic 
manifold at 
the critical dimension $d=2$ in a periodic potential, using a perturbative 
Functional Renormalization Group approach. Our aim is to describe the effective
potential seen by the manifold below the roughening temperature
on large length scales. We obtain analytically a flow 
equation for the potential and surface tension of the manifold, valid at
all temperatures. On a length scale $L$, the renormalized 
potential is made up of a succession of quasi parabolic wells, matching 
onto one another in a singular region of width $\sim L^{-6/5}$ for large $L$.
We also obtain numerically the step energy as a function of temperature, and
relate our results to the existing experimental data on $^4$He. Finally, we sketch the
scenario expected for an arbitrary dimension $d<2$ and
examine the case of a non local elasticity which is realized
physically for the contact line.
\end{abstract}

\vskip 0.5cm

\noi LPTENS preprint 99/XX

\vskip 0.5cm

\noi Electronic addresses : 
anusha@spec.saclay.cea.fr ;
bouchaud@spec.saclay.cea.fr


\section{Introduction}
The roughening transition has been studied in great detail, both theoretically
and experimentally \cite{Nozieres,BGR}. Direct analogies with the 
(two dimensional) 
$XY$-model or the Coulomb gas furthermore make this problem particularly 
enticing \cite{mappings}. More recently, the role  of disorder on the 
roughening transition or on the properties of the $XY$  model, 
has attracted considerable interest \cite{Cardy,BG,OS,CLD,EN}. In particular, 
replica calculations  and Functional Renormalization Group ({\sc frg}) methods have 
been applied to this problem, with sometimes conflicting results 
\cite{Comment}. In this paper, we wish to  reconsider the 
problem of the roughening transition in the absence of disorder, from a
{\sc frg} point of view, where the flow is not {\it a priori} 
projected onto the first harmonic of the periodic potential. Within a local 
renormalization scheme, we establish exact
equations for the evolution of the full periodic potential $V(\p)$, and the
surface tension $\ga$ with the length scale $L=e^\ell$, which we analyze 
both numerically and analytically, in the low temperature phase. 
If we start  with a sinusoidal periodic potential,  
the shape of the fixed point potential $V^*(\p)$ evolves to a nearly 
parabolic shape with matching points becoming more and more singular as the
length scale increases. 
The nature of the singularity  is investigated in detail
close to the fixed point, that is for small values of the 
rescaled temperature $\ds \ol T = \f{T}{2\pi \ga \l^2}$
where $\l$ is the periodicity of the potential and $\ga$ the elastic
stiffness. We find that the
width $\Delta \varphi$ of the singular region scales as $L^{-3g(\ol T)/5}$, where 
$g(\ol T)$ governs the scaling of the surface stiffness with the length
scale according to  $\ga(L)\sim L^{g(\ol T)}$. The exponent $g(\ol T)$ 
tends towards $2$  with negative  corrections which we calculate, 
when $\ol T$ goes to $0$ (i.e. for $L \to \infty$). 

The paper is organized as follows. In section $\mb 2$, we introduce the model:
we outline  the calculations involved and discuss the differences with the 
approach of Nozi\`eres and Gallet ({\sc ng}), and briefly
examine the problem for $d<2$. We then explain
in section $\mb 3$, by a mean field argument the origin of the singularity
that develops during the renormalization flow. In section $\mb 4$, we present
a scaling form for the renormalized potential, around its maxima and 
close to the fixed point, which accounts for the nature of the singularity.
Using our renormalization group flow, we compute in section $\mb 5$ the step
energy as a function of temperature. Finally, in section $\mb 6$, we look 
at the 
case of a contact line in a  periodic 
potential, as this is a physical realization of a non local elastic stiffness.

\section{Model and functional renormalization group}

We consider an elastic interface whose height fluctuations are described  by a
profile $\P(x)$, where $x$ is a $d$-dimensional vector, in the presence of a 
deterministic periodic potential $V$. Supposing that the slope of the
interface is everywhere small, the energy of the system is:

\be\label{energie}
H[\P]=\f{\ga}{2}\int d^dx\ (\na \P(x))^2 +  
\ds \int  d^dx\ V\biggl(\f{\P(x)}{\l}\biggr)
\ee 

\noi where $\ga$ is the elastic stiffness and $\l$ the periodicity of the 
potential. In the absence of periodic potential, the height fluctuations 
of the surface on a length scale $L$ scale as $L^{2-d}$. For $d>2$, the
interface is therefore always flat.
For the critical dimension $d=2$, the interface is rough only if the 
temperature exceeds a certain critical temperature $T_R$.
When the potential $V$ is harmonic,
this model is the continuous version of the Sine-Gordon model.
{\sc ng} have studied the statics of this problem using a 
two-parameter renormalization group scheme, and have written flow 
equations for $\ga(L)$ and the amplitude $v_o(L)$ of the periodic potential.
They suppose that during the flow, $v_o$ remains small 
compared with the temperature and neglect all higher harmonics of the 
potential. Correspondingly, within this procedure, the 
renormalization scheme ceases to be valid when $v_o$ becomes of the order 
of the temperature. In the low temperature `flat' phase, this occurs 
after a finite
renormalization since $v_o$ grows with distance.

In our calculation, we consider a general periodic function with the only 
constraint
that it should be sufficiently smooth 
(we shall explain this more quantitatively in the following). Since we re-sum 
the whole perturbation expansion in $v_o/T$, there is however no 
constraint on the amplitude of the potential, and the renormalization procedure
can be carried on any length scale without interruption. The relevant coupling constant appears to be $v_o/\gamma$ rather than $v_o/T$. 
During the  renormalization flow, we keep track of the whole function $V(\p)$
instead of projecting onto the first harmonic, so that we have a more
quantitative knowledge of the behaviour of the potential for low temperatures.

Technically, we proceed by considering the partition function:

\be\label{partition1}
\ds Z=\int d\P(x) \ e^{-\b H[\P(x)]}
\ee

\noi We perform the renormalization procedure by splitting the field $\P$ into
a slowly-varying and a rapidly-varying part as:

\be\label{split}
\P(x)=\P^<(x)+\P^>(x)
\ee 

\noi The  Fourier modes  $k$ of $\P^<$ are such that  
$0 \leq |k| \leq |\L|/s$, and those of $\P^>$,  such that 
$|\L|/s \leq |k| \leq |\L|$, where $s=e^{d\ell}$, $\ds |\L|$
being a high momentum cut-off, of the order of $1/a$, where $a$ is the
lattice spacing. We integrate over  the fast modes in the partition function 
and retain
only the terms that renormalize the gradient term and the potential term.
The other terms that are generated are discarded as irrelevant. Within
this renormalization scheme, our calculation is exact. After some algebra 
detailed in appendix ${\mb A}$, we obtain a set of flow 
equations  for $d=2$, for 
$\ds \ol V=\f{V}{\ga \l^2 |\L|^2}$ and  $\ds \ol T=\f{T}{2\pi \ga \l^2 }$, 
the rescaled potential and temperature (note that $\ol V$ and $\ol T$ are 
dimensionless):

\be\label{flow}
\ba{l}
\ds \f{d \ol V}{d\ell}=\ds (2-g)\ol V - \pi \f{{\ol V'}^2}{(1+{\ol V}'')} +
                       \f{\ol T}{2} \ln(1+{\ol V}'')

\\ \\

\ds \f{d \ga}{d\ell}=  g \ga

\\ \\

\ds \f{d \ol T}{d\ell}= - g \ol T

\ea
\ee

\noi where $g$ is given by:

\be
\label{Exp_g}
g=4\pi  \  \ds \int_0^{1} d\p \
\f{\ds{\ol V}'^2(\p){\ol V}'''^2(\p)}
{\biggl(1+{\ol V}''(\p)\biggr)^5}
+ \f{\ol T}{4}  \int_0^{1} d\p \ 
\f{{\ol V}'''^2(\p)}{\biggl(1+{\ol V}''(\p)\biggr)^4}\ee

\noi These equations call for some comments. 

\begin{figure}
\centerline{\hbox{\epsfig{figure=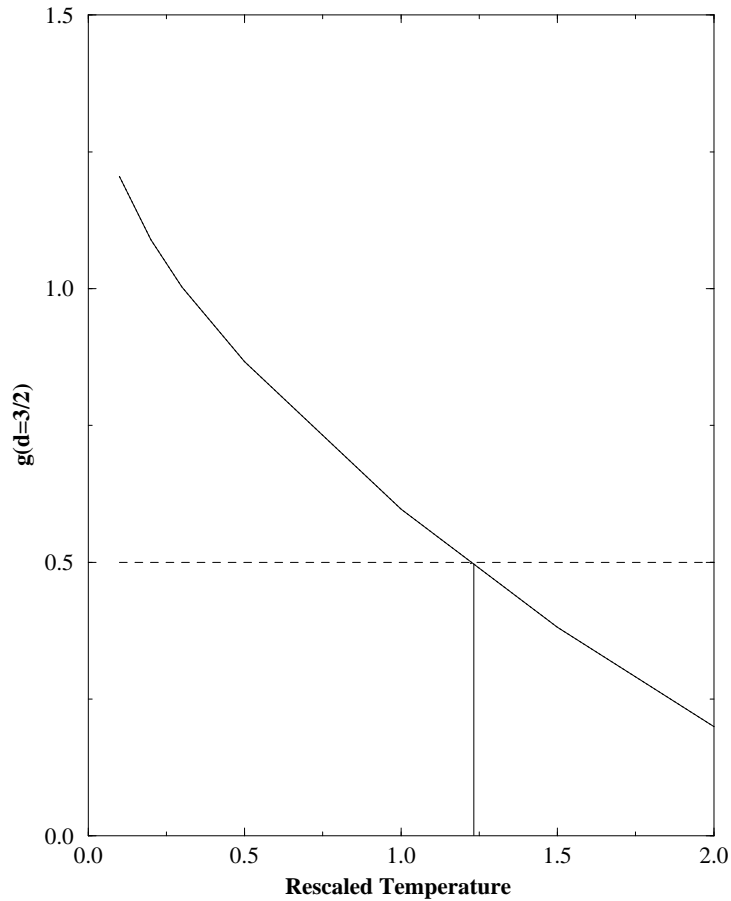,width=8cm}}}
\vskip 0.8cm
\caption{$g$ as a function of the rescaled temperature $\ol T$ for d=3/2. 
The point where $g=2-d=1/2$ determines the transition temperature.}
\label{fig1}
\end{figure}

\begin{itemize}

\item The relevant perturbative parameter appears to be $\ol V$, rather than $V/T$. In the limit $\ol V \ll 1$, and in the case where the 
potential is
purely harmonic (i.e. $V(\p)=v_o \cos(2\pi\p)$), the {\sc rg} equations read:

\be\label{NG}
\ba{l}
\ds \f{du_o}{d\ell}= \biggl(2- \f{\pi T}{\ga \l ^2}\biggr)u_o

\\ \\

\ds \f{d\ga}{d\ell}= 2 \pi^4 \left(\f{2 \pi T}{\ga \l^2}\right)
\f{u_o^2}{\ga^2 \l^4}
\ea
\ee

\noi where $u_o=v_o/|\L|^2$. The first equation is trivial and 
identical to the one in {\sc ng}, and immediately leads to the value of the 
roughening transition temperature: $T_R=2\ga_\infty \lambda^2/\pi$, where
$\ga_\infty$ is the renormalized value of $\ga$. The second is close to,
but different from the one obtained in the particular renormalization scheme 
used by {\sc ng}: near the critical temperature $T_R$,
the coefficient between parenthesis is equal to $4$ in our case and to $0.4$ 
according to {\sc ng}. 

\item The renormalization of the surface tension, as measured by $g$, 
is always 
{\it positive}. One can check that, as has been pointed out by {\sc ng}, 
if the initial potential
is parabolic (i.e. $V(\phi)=v_0 \phi^2$), then the coefficient $g$ vanishes
identically, and there is no renormalization of the surface tension. This is
indeed expected since in this (quadratic) case, all modes are decoupled.

\item The flow equations only make sense if $\ol V''>-1$. We have checked 
numerically that if this condition is satisfied at the beginning, it prevails 
throughout the flow. On the other hand, if the initial potential is so steep
that this condition is violated, the perturbative calculation is meaningless.
This comes from the fact that metastable states, where the surface zig-zags
between nearby minima of the potential, appear at the smallest length scales. 
In this respect, it is useful to note that the last term of the flow equation
on $\ol V$ comes from the integration of the Gaussian fluctuations of the
fast field around the slow field. The condition $\ol V''>-1$ is a stability 
condition for these fast modes. If the unrenormalized potential is 
harmonic (i.e. $V(\p)=v_o \cos(2\pi\p)$), and the unrenormalized surface
tension given by $\ga_o$, then this condition reads 
$\ds \f{v_o}{\ga} \biggl(\f{2\pi}{\l|\L|}\biggr)^2 < 1$, which 
simplifies to $\ds \f{v_o}{\ga} < 1$ in the case where $\l = a$. If the
initial value of the potential is too large, one actually expects the transition to become first order (but see \cite{Nozieres}). Actually, a variational calculation indeed predicts the transition to become first order when 
$\ds \f{v_o}{\ga} \geq 1$ \cite{Saito}.

\item The fundamentally new term in the above equation is the second 
term, proportional to $\ol V'^2$, and independent of temperature. This
term leads to the appearance of singularities in the flow equation: up to
second order in $\ol V$, this equation is close to the Burgers' equation
(see below) for which it is well known that shocks develop in time. The fact 
that this term survives even in the zero temperature limit is at first sight 
strange, since one could argue that for $T=0^+$, there are no longer any 
thermal
fluctuations, and thus no renormalization. This argument is not correct because
we are computing a {\it partition function}, thereby implicitly assuming that
the infinite time limit is taken {\it before} the zero temperature limit. 
Such a non trivial renormalization has also been found in the context of 
pinned manifolds \cite{BF,BBM}, and can be understood very simply using 
a mean-field approximation, which we detail in the next section.

\item For completeness let us consider  the case $1<d<2$. By simple scaling
arguments, we can see that for $d<1$ the interface is always rough. For $1<d<2$,
we have a roughening transition between a flat phase and a rough phase.
In this case, we allow $\P$ to renormalize and suppose that $\l$ renormalizes 
in the same way according to:

\be\label{ren_lambda}
\ds \f{d\l}{d\ell}=\z \l
\ee

\noi The other flow equations in terms of the rescaled parameters 
$\ds \ol V=\f{V}{\ga \l^2 |\L|^2}$ and  
$\ds \ol T=\f{K_d |\L|^{d-2}T}{\ga \l^2 }$  
(with $K_d=\ds S_d/(2\pi)^d$ where $S_d$ is the $d$-dimensional sphere) 
now read:

\be\label{flow_d}
\ba{l}
\ds \f{d \ol V}{d\ell}=\ds (d-g_d-2\z)\ol V - \pi \f{{\ol V'}^2}{(1+{\ol V}'')} +
                       \f{\ol T}{2} \ln(1+{\ol V}'')

\\ \\

\ds \f{d \ga}{d\ell}=  (g_d+d-2) \ga

\\ \\

\ds \f{d \ol T}{d\ell}= - (g_d+d-2+2\z) \ol T

\ea
\ee

\noi where $g_d$ is given by:

\be
\label{Exp_g_d}
\ds g_d=\f{2}{d} g,
\ee

\noi with $g$ given by equation (\ref{Exp_g}) above. These equations have a non trivial fixed point for $g_d=2-d$ and $\z=0$.
This corresponds to a rescaled temperature $T_R$ and a renormalized rescaled
potential $\ol V$ such that equations (\ref{flow_d}) 
and (\ref{Exp_g_d}) are satisfied.
For $1<d<2$, we obtain $T_R$ numerically by proceeding as follows:
we self-consistently solve the differential equation on  $\ol V$ obtained by putting 
$\ds \f{d\ol V}{d\ell}=0$ for different fixed rescaled temperatures $\ol T$,
imposing that $g$ is given by equation (\ref{Exp_g}). This enables us to plot
$g$ as a function of $\ol T$. The rescaled temperature $\ol T_R$ corresponding 
to the  transition temperature is such that $g_d=2-d$. In the figure (\ref{fig1}), we 
have plotted the result for $d=3/2$, with $\gamma_o=1$. In that case, $\ol T \simeq 1.2$.

\end{itemize}

\section{Mean field analysis and effective potential}

In this section, we show on a simplified mean field version of
the model how the non linear term $V'^2$ arises in the flow equation of the 
potential. Using a discrete formulation of the problem and replacing the
local elasticity modeled by the surface tension term by a coupling to all
neighbours, we can rewrite the energy as:

\be\label{H_mf}
\ds H_{\it mf}(\{\P\}_i,\ol \P)=\f{\ga}{2a} \sum_i^N (\P_i-\ol \P)^2
+a \sum_i^N V(\P_i)
\ee

\noi where $\ol \P$ is the center of mass of the system, $a$ the lattice
spacing and $L=Na$. Implementing the constraint $\ol \P =\ds 1/N \sum \P_i$
by means of a Lagrange multiplier in the partition function, we have:

\be\label{Z_mf}
\ds Z[\ol \P] = \ds \int d\eta \int \prod_i d\P_i \
e^{-\ds \b H_{\it mf}(\{\P\}_i,\ol \P)-
\eta \biggl(N\ol \P-\ds \sum_i \P_i\biggr)}
\ee

\noi which can also be expressed as:

\be\label{Zfact}
\ds Z[\ol \P] = 
\int d\eta \ e^{\ds N \log z(\ol \P,\eta)+\ds N \f{\eta ^2 a}{2\b \ga}}
\ee

\noi where

\be\label{zo}
\ds z(\ol \P,\eta)= 
\int d\P \ 
e^{\ds -\f{\b \ga}{2a} \biggl(\ol \P - \P+ \ds \f{\eta a}{\b \ga}\biggr)^2
-\b aV(\P)}=
\ds z\biggl(\ds \ol \P+\f{\eta a}{\b \ga}\biggr)
\ee

\noi We are left with a simpler problem since we now have a one-body problem.
We introduce the auxiliary partition function $z_R(\Psi, \tau)$ defined as:

\be\label{diff}
\ds z_R(\Psi, \tau)=\ds \sqrt \f{\b \ga}{2\pi a \tau}
\ds \int d\P \ e^{\ds -\f{\b \ga}{2a} \f{(\Psi-\P)^2}{\tau}
-\b a V(\P)}
\ee

\noi Up to a multiplicative constant, one has 
$\ds z\biggl(\ds \ol \P+\f{\eta a}
{\b \ga}\biggr)=
\ds z_R \biggl(\ds \ol \P + \f{\eta a}{\b \ga}, \tau=1\biggr)$,
 where $z_R(\Psi, \tau)$ verifies the diffusion equation:

\be\label{equadiff}
\ds \f{\pa z_R}{\pa \tau}= \f{a}{\b \ga} \f{\pa ^2 z_R}{\pa \Psi ^2}
\ee
with an initial condition given by:

\be\label{init}
\ds z_R(\Psi, \tau=0)=  e^{-\ds \b aV(\Psi)} 
\ee

\noi Defining now the effective pinning potential $V_R$ as:

\be\label{V_R}
aV_R\biggl(\ol \P + \ds \f{\eta a}{\b \ga},\tau\biggr)=
-T\log z_R \biggl(\ol \P + \f{\eta a}{\b \ga}, \tau\biggr)
\ee

\noi we can then easily show that $V_R$ is the Hopf-Cole
solution of the non linear Burgers' equation \cite{BBM}:

\be\label{diffVren}
\ds \f{\pa V_R}{\pa (\tau a)}=\f{T}{\ga} \f{\pa ^2 V_R}{\pa \Psi ^2}
- \f{a}{\ga} \biggl(\f{\pa V_R}{\pa \Psi}\biggr)^2
\ee

\noi where the temperature independent non linear term $V_R'^2$ indeed appears. 

It is easy to show that when $N \to \infty$ or $T \to 0$, the original 
partition function can be solved by a saddle point method, leading after a change
of variables to an  effective potential per unit length:

\be\label{V_eff_final}
\ds V_{\it eff}(\ol \P)=V_R (u,\tau=1)-\ds \f{\ga}{2a^2}(\ol \P-u)^2
\ee

\noi where $u$ is given by:

\be\label{eta*}
\ds \f{\ga}{2a}(u-\ol \P)= \f{\pa V_R}{\pa \p}(\p,\tau=1) \ds \biggr|_{\p=u} 
\ee

\noi Now, by changing $V_R$ to $-V_R$, one can see that $-V_{\it eff}$ can also
be written as the solution of a Burgers' equation. It is known from results on the Burgers' equation, that, with `time' $\tau$, 
the effective potential $V_R$ develops shocks, smoothed out at finite temperature, 
between which it has a parabolic shape. The appearance of singularities is
due to the non linear term in the partial differential equation, {\it which 
indeed
survives in the limit} $T=0$. It is interesting to see how this `toy'
renormalization group captures some important features of the full scheme, 
such as the one shown above for a non disordered potential.

\section{Analysis for small $\ol T$}

In this section, we go back to the model introduced in section $\mb 2$ and
analyze the nature of  $\ol V$ close to the low temperature fixed point, 
that is for small values of the rescaled temperature ${\ol T}$. Since $g >0$ 
in the
low temperature phase, this corresponds to the large scale structure of the
renormalized potential for all temperatures $T<T_R$.

Expanding ${\ol V}$ around one of its 
minima as ${\ol V}(\p)={\ol V}_{m}+\f{1}{2}\ka (\p-\p^*)^2$, and replacing 
${\ol V}$ in the flow equation(\ref{flow}), we have

\be\label{flow_min}
\ba{l}
\ds \f{d \ol V_m}{d\ell}=\ds (2-g)\ol V_m + \f{\ol T}{2} \ln(1+\ka)

\\ \\

\ds \f{d \ka}{d\ell}=\ds (2-g)\ka -2\pi \f{\ka ^2}{1+\ka}

\ea
\ee 

\noi One can actually check that a parabolic shape for $\ol V$ is exactly
preserved by the renormalization flow. However, since the potential has
to be periodic, these parabolas should match periodically around each 
maximum that is for $\p-\p^*=0, 1, 2,...$. 
The region of the maximum is therefore expected  to be singular. 
To investigate the nature of the renormalized periodic potential 
around its  maximum value, we will thus make a scaling ansatz on ${\ol V}''$ 
for small $\ol T$. For our perturbative calculation to be valid, we expect
$\ol V''(0)$ to be $> -1$. Now since we expect a singularity to 
develop as  $\ol T$ goes to $0$, it is probable (and actually 
self-consistently checked) that $\ol V''(0)$ should 
tend towards $-1$. As $\ol T$ goes to $0$, we thus make the scaling ansatz:

\be\label{ansatz_V''}
\ds 1+{\ol V}''(\p)=\ds {\ol T}^{\d}{\cal F}'
\biggl(\f{\p}{{\ol T}^{\a}}\biggr)
\ee

\noi where ${\cal F}'(0) > 0$. This means that the width of the singular
region behaves as $\Delta \p \sim {\ol T}^{\a}$. Hence, in the scaling region:

\be\label{ansatz_V'}
\ds {\ol V}'(\p)=\ds -\p+{\ol T}^{\d+\a}{\cal F}
\biggl(\f{\p}{{\ol T}^{\a}}\biggr)
\ee
with ${\cal F}(0) = 0$ to ensure that $\p=0$ is a maximum of $\ol V$. Integrating once more the above equation, one finds:

\be\label{ansatz_V}
\ds {\ol V}^{}(\p)=\ds {\ol V}_M-\f{\p^2}{2}+
{\ol T}^{\d+2\a}{\cal G}\biggl(\f{\p}{{\ol T}^{\a}}\biggr)
\ee

\noi with ${\cal G}'= {\cal F}$. Replacing this last equation in the flow 
equation for $\ol V$, we obtain:

\be\label{V_max}
\f{d \ol V_M}{d\ell}=(2-g){\ol V}_M + \f{\ol T}{2}\log{\ol T}^{\d}
\ee

\noi Suppose that equations (\ref{flow_min}) have a fixed point as $\ol T$
goes to zero, and that close to the fixed point one can neglect the 
left hand side of these equations. This leads to the relation
$\ka=(2-g)/(2\pi-2+g)$. Supposing moreover that the parabolic solution 
extends almost over a whole period and that the correction brought about by 
the rounding off of the singularity around the maxima of the potential is 
negligible, we also have

\be\label{Eq_1}
\ds ({\ol V}_M-{\ol V}_m) \simeq \f{\ka}{2}  
\ds = \f{2-g}{4\pi-4+2g}
\ee

\noi Now, subtracting equation (\ref{V_max}) from equation (\ref{flow_min}), 
we find in the limit $\ol T \to 0$:

\be\label{Eq_2}
\ds \f{d}{d\ell}({\ol V}_M-{\ol V}_m)=
(2-g)({\ol V}_M-{\ol V}_m) + \f{\d}{2} \ol T \log {\ol T}
\ee

\noi Combining equations (\ref{Eq_1}) and (\ref{Eq_2}), we find that for the 
previous  equation to have a fixed point as $\ol T \to 0$, $g \to  2$
with negative corrections as:

\be\label{2-g}
(2-g) \simeq \ds \sqrt{2\pi \d{\ol T}\log \f{1}{\ol T}}
\ee

\noi This result is independent of the way we calculate $g$, the correction
to the surface tension. In particular, it shows that at zero temperature,
the surface tension diverges as $(L/a)^2$, where $L$ is the size of 
the system.

We can deduce an equation satisfied by ${\cal F}'$, by plugging the ansatz 
for the derivatives of $\ol V$ in the flow equation for $\ol V'$:

\be\label{eq_renV'}
\ds \f{d\ol V'}{d\ell}= (2-g-2\pi)\ol V' + 2\pi \f{\ol V'}{1+\ol V'} 
+ \pi {\ol V'}^2 \f{\ol V'''}{(1+\ol V'')^2}
+\f{\ol T}{2} \f{\ol V'''}{1+\ol V''}
\ee

\noi Close to the fixed point, we again suppose that to leading order in 
$\ol T$,
$\ds \f{d\ol V}{d\ell}=0$ in the above equation.  Plugging in the ansatz for
${\ol V}'$ and ${\ol V}''$, and considering the leading term 
in ${\ol T}$, we get to lowest order in $\ol T$:

\be\label{eq_calF}
-\ds \f{2\pi u}{{\cal F}'(u)} \ {\ol T}^{\a-\d} 
+\f{\pi u^2{\cal F}''(u)}{{\cal F}'(u)} \ {\ol T}^{\a-\d} 
+\f{{\cal F}''(u)}{2{\cal F}'(u)} \ {\ol T}^{1-\a} =0
\ee

\noi We can show that  necessarily $\a-\d=1-\a$. Indeed, if  $1-\a < \a-\d$, 
${\cal F}''$ would be equal to zero while  if $1-\a > \a-\d$, ${\cal F}'(0)$ 
would be equal to zero, both alternatives being thus impossible. 
Hence, equation (\ref{eq_calF}) can be rewritten as

\be\label{eq_calF2}
\ds \f{1}{2} \f{d}{du} \log {\cal F}' 
- \pi \f{d}{du} \biggl(\f{u^2}{{\cal F}'}\biggr) =0
\ee

\noi which yields after integration:

\be\label{Eq_calF3}
\ds {\cal F}'(u)\log \biggl( \f{{\cal F}'(u)}{{\cal F}'(0)}\biggr)=2\pi u^2
\ee

\noi At this stage, we can note that the exponent relation 

\be\label{Rel_exp1}
2\a-\d=1
\ee

\noi is independent of the scheme used to calculate of $g$ (see Appendix).

In the rest of this section, we calculate the exponents $\a$ and $\ga$, and
using $g(\ol T \to 0)=2$, we also obtain ${\cal F}'(0)$. These results now 
somewhat depend on the precise renormalization scheme we use to calculate the correction
$g$ to the surface tension. Replacing the derivatives
of $\ol V$ by their expressions in terms of ${\cal F}'$ and ${\cal F}''$, 
in equation (\ref{Exp_g}), and changing variables 
from $\p$ to  $\ds u=\f{\p}{{\ol T}^{\a}}$, we get to lowest order in $\ol T$:

\be\label{eq_g(T)}
\ds g(\ol T) \simeq {\ol T}^{\a-3\d} \ 8\pi 
\int_0^{\infty} du \ u^2 \f{{\cal F}''^2(u)}{{\cal F}'^5(u)}
+ {\ol T}^{1-\a-2\d} \ \f{1}{2} \int_0^{\infty} du \
\f{{\cal F}''^2(u)}{{\cal F}'^4(u)}
\ee

\noi From the exponent relation $2\a-\d=1$ derived previously, and the
fact that $g(\ol T \to 0)$ is finite, we have another exponent relation 
$\a=3\d$,  so that
$\a=3/5$ and $\d=1/5$. We can also show that $g(\ol T \to 0)$ can be 
expressed in terms 
of ${\cal F}'(0)$. From expression (\ref{Eq_calF3}), one can see that 
${\cal F}'$ is a strictly 
increasing function on $[0,\infty]$ taking its values in $[e,\infty]$,
and so we can change  variables from  ${\cal F}'$ to its inverse function.
Defining a new variable $x$ as 

\be\label{cgt_var}
\ds x = \f{e {\cal F}'(u)}{{\cal F}'(0)}
\ee

\noi we have: 

\be\label{eq_u(x)}
\ds u= \biggl( \f{{\cal F}'(0)}{2\pi e} \biggr)^{1/2} (x\log (x))^{1/2}
\ee

\noi and

\be\label{eq_du/dx}
\ds \f{du}{dx}= \f{1}{2} \f{\log(x)}{(x\log(x))^{1/2}}
\biggl( \f{2\pi e}{{\cal F}'(0)} \biggr)^{1/2}
\ee
 
\noi We can now express $g(\ol T \to 0)$ in terms of an integral over $x$ from 
$e$ to $\infty$ as

\be\label{eq_g(0)}
\ds g(0)= \f{2}{\pi ^2} \biggl(\f{2\pi e}{{\cal F}'(0)} \biggr)^{5/2}
\biggl\{ \int_{e}^{\infty} dx \ \f{(x \log (x/e))^{3/2}}{x^5\log (x)} +
\int_{e}^{\infty} dx \ \f{(x \log (x/e))^{1/2}}{x^4\log (x)} \biggr\}
\ee
Hence, using the fact that $g(\ol T \to 0)=2$, we finally find the constant
${\cal F}'(0) \simeq 1.38$.

\section{Step energy as a function of temperature}

\begin{figure}
\centerline{\hbox{\epsfig{figure=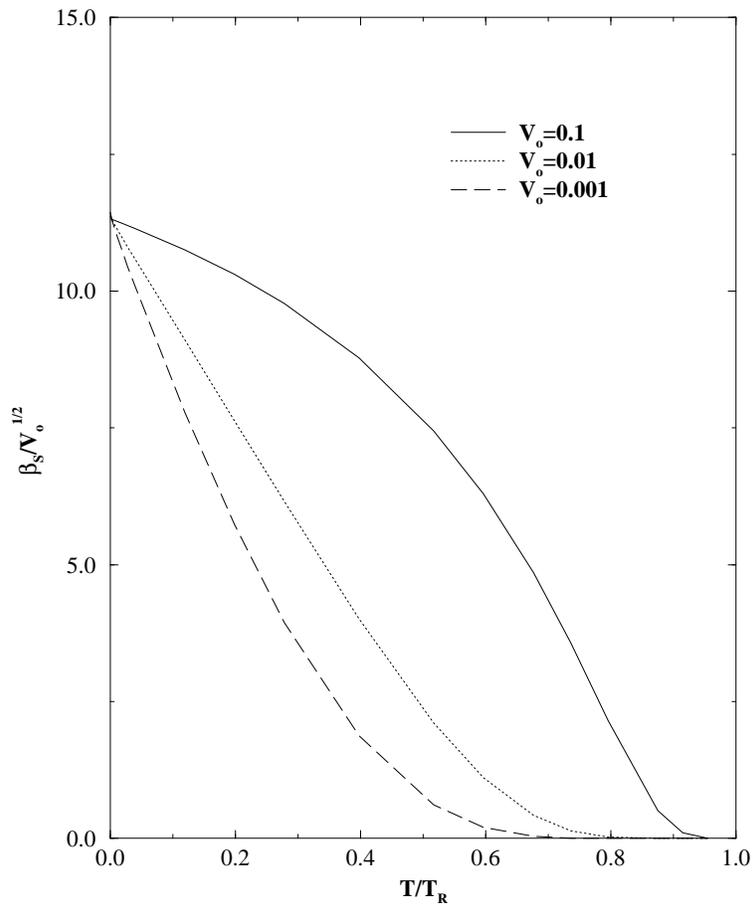,width=8cm}}}
\vskip 0.8cm
\caption{$\beta_S/\protect\sqrt{\ol v_o}$ as a function of the rescaled temperature for three different values of the bare periodic potential $\ol v_o$, with $\gamma_o=1$.}
\label{figure2}
\end{figure}

From physical considerations we know that below the roughnening temperature,
the interface grows by forming terraces. An important quantity governing the
kinetics of growth is therefore the step energy. The width $\xi$ of a step
and its energy per unit length $\beta_S$ can be obtained by comparing the 
elastic energy and the potential energy of a profile $\Phi(x)$ which changes 
by one period 
over the length $\xi$. Requiring that these two energies are of the same order
of magnitude leads to: $\ds \ga/\xi ^2 \sim v_o$ where $v_o$ is the
amplitude of the periodic potential, or $\ds \xi \sim \sqrt {\f{\ga}{v_o}}$,
and a step energy which scales as $\ds \b_S \propto \sqrt{{v_o}{\ga}}$. 
Since a step 
profile include Fourier modes such that $\xi^{-1} < k < |\L|$, it is natural
to use in the above equations the values of $\ga$ and $v_o$ calculated for 
the length $L = a e^{\ell} = \xi$. Since $\xi(L) \sim L/\sqrt{\ol{v}_o(L)}$, one sees 
that this corresponds to
stopping the renormalization procedure when $\ol{v}_o(L) \sim 1$. We have 
integrated numerically the {\sc rg} flow, starting from $\ga=1$ and from 
harmonic potentials of various amplitudes $v_o \ll 1$, and stopping for an
arbitrary value $\ol V$, chosen here to be $\ol v_c=0.4$. \footnote{Other values of $\ol v_c$ would not change the qualitative features reported below, provided 
$\ol v_c$ is not too large.} The resulting step energy as a function of temperature is 
plotted in Figure (\ref{figure2}). For $T$ close to $T_R$, one finds that $\xi$
diverges as $\ds e^{1/\sqrt{T_R-T}}$, as it should since our {\sc rg} flow 
essentially boils down to the standard one \cite{Nozieres}. For small temperatures, however, we find that
$\beta_S$ tends to a finite value with a linear slope in temperature. 
This slope is seen to decrease as the initial amplitude of the potential $\ol v_o$ increases. For $\ol v_o=0.01$, $\beta_S$ decreases by $\sim 30 \%$ when $T$ increases 
from $0$ to $0.25\ T_R$. This decrease falls to $\sim 10 \%$ for 
$\ol v_o=0.1$.

Experiments on Helium 4, on the other hand, have established that the step 
energy does only depend very weakly on temperature at small temperature, by not more than $5\%$ when the temperatures varies from 
$0.05 \ T_R$ to $0.25 \ T_R$. This suggests that the initial amplitude 
of the potential is of the same order as $\gamma_o$: in this case, 
the width of the step is of order $a$, and the bare parameters are not 
renormalized except possibly very close to $T_R$. The conclusion that 
experiments must be in the regime $\ol v_o \sim 1$ is in 
agreement with \cite{BGR}, where $\ol v_o$ is called $t_c$ (up to a numerical prefactor); $v_o/\gamma_o$ was estimated to be $\sim 0.05$. Since our {\sc rg} flow is different from the one 
obtained by Nozi\`eres and Gallet, the values of the
physical parameters obtained by a fit of our theory to the experiments will actually differ.

\section{Case of the contact line}

In this section, we repeat the previous analysis for the case of a contact 
line on a periodic substrate \cite{DG,JDG}. 
The roughness of a contact line on a  disordered substrate, 
at zero temperature, has been studied analytically and compared
with the experimental predictions for the case of superfluid helium on a 
disordered cesium substrate, where the disorder arises from randomly 
distributed wettable heterogeneities which are oxydized areas of the substrate
\cite{Rolley,HM}. 
A physical realization of the theoretical situation we consider 
here could be achieved by preparing a substrate with equally spaced oxydized 
lines which would act as periodic pinning grooves. In this case the critical 
dimension is $d=1$. We denote by $\P$ the position of the line with respect
to a mean position. The energy of the system is the sum of an elastic term
and a potential term given by:

\be\label{energie_ligne}
\ds H[\P]=\f{\ga}{2} \int \ \f{dk}{2\pi} |k||\P(k)|^2+
\int_0^L dx \ V(\P(x))
\ee

\noi where $L$ is the length of the substrate and $\ga$ the stiffness. 

The renormalization procedure is carried as before except that the propagator
is now given by $\ds G(k)=\f{1}{\b \ga |k|}$. Moreover the renormalization
of the stiffness now only  comes from the scale change leading to the much
simpler flow equation for $\ga$:

\be\label{ren_gamma}
\ds \f{d\ga}{d\ell}=\ga
\ee

\noi Defining as before the rescaled parameters $\ol V$ and $\ol T$ with
$\ds \ol V=\f{V}{\ga \l^2 |\L|}$ and $\ds \ol T=\f{2T}{\ga \l^2}$, the flow
equations for $\ol V$ and $\ol T$ read:

\be\label{ren_V}
\ds \f{d\ol V}{d\ell}= \ol V -\f{\ol V'^2}{1+\ol V''}
+\f{\ol T}{2}\log(1+\ol V'')
\ee

\noi and

\be\label{ren_T}
\ds \f{d\ol T}{d\ell}=-\ol T
\ee

\noi During the flow, $\ol T$ flows to zero and
the renormalized  rescaled potential $\ol V$ develops  shocks between which 
it has a parabolic shape. We characterize the singularities that develop 
around  the maxima of $\ol V$ by the following scaling ansatz:

\be\label{scale_ligne1}
\ds 1+\ol V''(\p)= \ds T^{\d} \ds e^{-\f{A}{\ol T}} 
{\cal F}'  \biggl(\f{\p}{T^{\a} \ds e^{-\f{B}{\ol T}}}\biggr)
\ee

\noi where ${\cal F}'(0) > 0$ and ${\cal F}(0) = 0$. 
Putting $\ds u=\f{\p}{{\ol T}^{\a} \ds e^{ -\f{B}{\ol T}}}$, this implies that
for $u \sim 1$,

\be\label{scale_ligne2}
\ds {\ol V}'(u)=\ds -{\ol T}^{\a} \ds e^{ -\f{B}{\ol T}} u
+{\ol T}^{\d+\a} \ds e^{ -\f{A+B}{\ol T}} {\cal F}(u)
\ee

\noi and 

\be\label{scale_ligne3}
\ds {\ol V}(u)=\ds {\ol V}_M-
{\ol T}^{2\a} \ds e^{ -\f{2B}{\ol T}} \f{u^2}{2} +
{\ol T}^{\d+2\a} \ds e^{ -\f{A+2B}{\ol T}} {\cal G}(u)
\ee

\noi with ${\cal G}'= {\cal F}$. Plugging the previous expressions into
the flow equation $\ol V'$:

\be\label{ligne_V'}
\ds \f{d\ol V'}{d\ell}= -\ol V' + 2\ol V' \f{\ol V''}{1+\ol V'} 
+ {\ol V'}^2 \f{\ol V'''}{(1+\ol V'')^2}
+\f{\ol T}{2} \f{\ol V'''}{1+\ol V''}
\ee

\noi and supposing that  $\ds \f{d\ol V'}{d\ell}=0$, to leading order as  
$\ol T$ goes to zero, we have to leading order in $\ol T$:

\be\label{eq_calFbis}
-\ds \f{2u}{{\cal F}'(u)} \ {\ol T}^{\a-\d} \ds e^{\f{A-B}{\ol T}}
+\f{u^2{\cal F}''(u)}{{\cal F}'^2(u)} \ {\ol T}^{\a-\d} 
\ds e^{\f{A-B}{\ol T}}
+\f{{\cal F}''(u)}{2{\cal F}'(u)} \ {\ol T}^{1-\a} \ds e^{\f{B}{\ol T}}
=0
\ee

\noi Since ${\cal F}'(0)>0$, we obtain a non trivial solution only if

\be\label{ligne_rel_exp}
\ds 2\a-\d=1 \quad {\rm and} \quad A-B=B
\ee

\noi and ${\cal F}'$ is again solution of equation  
(\ref{Eq_calF3}). We can obtain the values of the parameters $A$ and $B$ by considering separately the 
singular part and the regular part of the renormalized rescaled potential
$\ol V$ close to the fixed point. We expand  $\ol V$ around one of its minima
$\p^*$ as:

\be\label{ligne_exp_min}
\ds \ol V(\p)=\ol V_m + \f{\ka}{2}(\p-\p^*)^2
\ee

\noi and plug the resulting expression in the flow equation for $\ol V$.
This yields:

\be\label{ligne_flow_min}
\ba{ll}
\ds \f{dV_m}{d\ell} & = \ds V_m+\f{\ol T}{2}\log(1+\ka) 

\\ \\

\ds \f{d\ka}{d\ell} & = \ds \ka - \f{2\ka ^2}{1+\ka}
\ea
\ee

\noi We note that the fixed point value for $\ka$ is now finite as $\ol T$ 
goes to zero and is given by $\ka^*=1$. Similarly the flow equation 
for $\ol V(0)=\ol V_M$ is:

\be\label{ligne_flow_max}
\ds \f{d\ol V_M}{d\ell}=\ol V_M+\f{\ol T}{2} \log 
\biggl(T^{\d}\ e^{-\f{A}{\ol T}}{\cal F}'(0)\biggr)
\ee

\noi Combining equations (\ref{ligne_flow_min}) and (\ref{ligne_flow_max}),
the flow equation for the amplitude of $\ol V$ is given to leading order
as $\ol T \to 0$ by:

\be\label{ligne_amp}
\ds \f{d}{d\ell}(\ol V_M- \ol V_m)=(\ol V_M- \ol V_m)-\f{A}{2}
\ee

\noi Now, we expect that the singularity brings but a small correction to
the parabolic part of the rescaled potential $\ol V$, so that
at the fixed point the amplitude of $\ol V$ is given by $\ds \f{\ka^*}{2}$,
leading to $A=1$. The value of the exponents $\a$ and $\d$ would require the 
analysis of subdominant terms. The conclusion of this section is that in the
case of the contact line, the width of the singular region of the renormalized
potential decreases exponentially with length scale: the potential quickly
becomes a succession of matched parabolas.

\section{Conclusion}

In this paper, we studied the problem of the thermal roughening 
transition using a {\sc frg} formalism. We have shown that below the roughening temperature, the
periodic potential on large length scales cannot be described by its lowest 
harmonic and that during the flow shocks are generated in the effective 
pinning potential. We expect that this result is more generally valid, and also holds in the case of a disordered pinning potential \cite{BF,BBM}. By performing  a resummation of our perturbation 
expansion, our results are in principle valid in the strong coupling regime, where the coupling constant is proportional to $V/\ga$ (rather than $V/T$). 
Correspondingly, we stop the renormalisation procedure not when $V(L) \sim T$
(as in {\sc ng}), but rather when $L$ reaches the size of the objects under
investigation (for example the width of the steps). By comparing our numerical result with the experimental determination of the step
energy of liquid Helium 4, we have concluded that the surface of Helium 4 crystals are such that the coupling to the lattice is of the same order of magnitude as the surface tension. This is in qualitative agreement with 
Balibar et al. \cite{BGR}, who estimate $v_o/\gamma_o \sim 0.05$. 

\section*{Acknowledgements}

We wish to thank Sebastien Balibar, T. Emig and M. M\'ezard 
for very interesting discussions.

\newpage

\appendix
\section{Derivation of the flow equations}
In this appendix, we sketch the procedure to obtain the flow equations
(\ref{flow}). We consider the partition function

\be\label{partition}
\ds Z =  \int d[\P]\  e^{\ds -\b {\cal H}(\P(x))}
\ee

\noi where ${\cal H}$ is the hamiltonian given by (\ref{energie}). We split 
the field $\P$ into a fast moving and a slow moving component and average
over the fast moving part. We can rewrite the partition function, up to a
multiplicative constant, as:

\be\label{part_moy} 
\ds Z = \int d[\P^<]\ e^{\ds -\b \int_< \f{d^dk}{(2\pi)^d} |k|^2|\P^<(k)|^2}
\ds \biggl< 
e^{\ds -\b \int d^dx \ V\biggl(\f{\P(x)}{\l}\biggr)}\biggr>_o 
\ee

\noi where $<...>_o$ represents the thermal average with respect to
the gaussian weight:

\be
e^{\ds -\b \int_> \f{d^dk}{(2\pi)^d} |k|^2|\P^>(k)|^2}
\ee

\noi In the rest of this section we denote $\ds \f{d^dk}{(2\pi)^d}$ by
$\t dk$.

\subsection{Renormalization of the periodic potential $V$}

We look for contributions to the potential $V$ resulting from
the above averaging, which are of the same form as the terms present 
in the hamiltonian before starting the renormalization procedure and 
which are   of order $d\ell$. These terms are represented by connected 
graphs and are all obtained by expanding the potential term  
with respect to $\P^>$  up to second order.
There are only two ways of obtaining such graphs of order $d\ell$:

$\bullet$
By contracting $p$ two-legged terms $\ds -\f{\b}{2} \ds \int d^dx \
\biggl(\f{\P^>(x)}{\l}\biggr)^2 
V''\biggl(\f{\P^<(x)}{\l}\biggr)$ 
with $1 \leq p \leq \infty$.
We must calculate:

\be\label{eq_ren1}
\ba{ll}
\ds \f{1}{p!} \biggl(-\f{\b}{2\l ^2}\biggr)^p 
& \ds \int \prod_{j=1}^p d^dx_j \  V''  \biggl(\f{\P^<(x_j)}{\l}\biggr)

\\ \\

& \ds \int \prod_{j=1}^p \t dk_j \t dk'_j  \ds \  
e^{i(k_1+k'_1)x_1...i(k_p+k'_p)x_p}
\biggl<\P^>(k_1)...\P^>(k'_p)\biggr>_o
\ea
\ee

\noi which gives after averaging over the fast modes:

\be\label{eq_ren2}
\ds \f{(-1)^p}{2p} \biggl(\f{1}{\ga \l^2}\biggr)^p   
\ds \int \prod_{j=1}^p d^dx_j   V''\biggl(\f{\P^<(x_j)}{\l}\biggr)
\ds \int \prod_{j=1}^p \t dk_j \ds \  
\f{\ds e^{ik_1(x_1-x_2)+...ik_p(x_p-x_1)}}{|k_1|^2...|k_p|^2} 
\ee

\noi In the above expression, the space dependence of $V''$ is slowly 
varying, and since the integral is dominated by the region where
the $x_j$'s are close to one another, we can with little error,
treat these terms as approximately equal to
$\ds V''\biggl(\f{\P^<(x_1)}{\l}\biggr)$. After integrating  over the
rest of the $x_j$'s and  summing up over  $p$, we are left with:

\be\label{eq_ren3}
-\b d\ell \  \f{{\cal K}_d |\L|^d T}{2} \ds \int d^dx \ \log 
\biggl(1+ \f{V''}{\ga \l^2 |\L|^2}\biggr)
\ee

$\bullet$
By contacting $2$ one-legged terms 
$-\b \ds \int d^dx \ 
\biggl(\f{\P^>(x)}{\l}\biggr)V'\biggl(\f{\P^<(x)}{\l}\biggr)$
with $p$ two-legged terms 
$\ds -\f{\b}{2} \ds \int d^dx \ \biggl(\f{\P^>(x)}{\l}\biggr)^2 
V''\biggl(\f{\P^<(x)}{\l}\biggr)$  
with $0 \leq p \leq \infty$.
To illustrate our method, we begin with $p=0$.
Expressing the fast modes in Fourier space, we have in discrete space:

\be\label{eq_ren4}
\ds \f{1}{2!} \b^2 \biggl(\f{1}{\l^2}\biggr)
\biggl(\f{a}{L}\biggr)^{2d}\ds \sum_{x} \sum_{y}
V'\biggl(\f{\P^<(x)}{\l}\biggr) V'\biggl(\f{\P^<(y)}{\l}\biggr)
\sum_{k} \sum_{k'} \ds e^{ikx+ik'y}
\biggl<\P^>(k)\P^>(k')\biggr>_o
\ee

\noi which gives after averaging over the fast modes:

\be\label{eq_ren5}
\ds  \f{\b}{2} \biggl(\f{1}{\ga \l^2}\biggr)  a^{2d} \sum_{x} \sum_{y}
V'\biggl(\f{\P^<(x)}{\l}\biggr) V'\biggl(\f{\P^<(y)}{\l}\biggr)
\f{1}{L} \sum_{k} \ds \f{e^{ik(x-y)}}{|k|^2}
\ee

\noi In the above expression, the main contribution comes from the part $x=y$,
while the part with $x \neq y$, has a phase which averages to almost zero.
Using the fact that $\ds |\L| = \f{2\pi}{a}$, the result is:

\be\label{eq_ren6}
\ds \b  d\ell  \f{K_d (2\pi)^d}{2} \int d^dx \
\f{{\ol V}'^2(\P^<(x)/\l)}{\ga \l^2 |\L|^2}
\ee

\noi Proceeding in a similar way for $1 \leq p \leq \infty$, we have

\be\label{eq_ren7}
\ba{ll}
\ds \b^2 & \ds  \biggl(\f{-\b}{2}\biggr)^p \f{1}{(p+2)!}  \f{(p+1)(p+2)}{2}  
\biggl(\f{1}{\l^2}\biggr)^{p+1}
\ds \biggl(\f{a}{L}\biggr)^{(p+2)d}

\\ \\

& \ds \sum_{x,y,x_1...x_p}
V'\biggl(\f{\P^<(x)}{\l}\biggr) V'\biggl(\f{\P^<(y)}{\l}\biggr)
V''\biggl(\f{\P^<(x_1)}{\l}\biggr)...V''\biggl(\f{\P^<(x_p)}{\l}\biggr)

\\ \\

& \quad \quad \ds \sum_{k,k',k_1...k'_p} \ds 
e^{ikx+ik'y+i(k_1+k'_1)x_1+...+i(k_p+k'_p)x_p}
\biggl<\P^>(k)\P^>(k')\P^>(k_1)...\P^>(k'_p)\biggr>_o
\ea
\ee

\noi which yields  after retaining the $x=y$ part and 
averaging over the fast modes: 

\be\label{eq_ren8}
\ba{ll}
\ds \f{\b}{2} (-1)^p a^d \biggl(\f{1}{\l^2}\biggr)^{p+1} 
\ds \int  \prod_{j=1}^p d^dx_j d^dx &
V'^{2}(\P^<(x))   \ds \prod_{j=1}^p V''(\P^<(x_j))

\\  \\

& \ds \int \prod_{j=1}^p \t dk_j \t dk \
\ds \f{e^{ik(x-x_1)+ik_1(x_1-x_2)...+ik_p(x_p-x)}}{|k|^2|k_1|^2...|k_p|^2}
\ea
\ee

\noi Treating the above expression as equation(\ref{eq_ren2}) we
are left with:

\be\label{eq_ren9}
\ds \b  d\ell \ (-1)^p  \f{K_d (2\pi)^d}{2} \int d^dx \
\biggl(\f{V''(\P^<(x)/\l)}{\ga \l^2 |\L|^2}\biggr)^p
\ee

\noi Summing up over $p$, we finally obtain:

\be\label{eq_ren10}
\ds \b  d\ell \  \f{K_d (2\pi)^d}{2} \int d^dx \
\f{V'^{2}(\P^<(x)/\l)}{\ga \l^2 |\L|^2}
\biggl(1+\f{V''(\P^<(x)/\l)}{\ga \l^2 |\L|^2}\biggr)
\ee

\noi Taking into account the rescaling of the potential term, and 
supposing we are in the flat phase so that $\P$ 
and $\l$ are not rescaled, we obtain for $d=2$:

\be\label{eq_ren11}
\ds \f{dV}{d\ell}=\ds 2V - \pi \
\f{\ds \biggl(\f{V'^{2}}{\ga \l^2 |\L|^2}\biggr)}
{\ds 1+\biggl(\f{V''}{\ga \l^2 |\L|^2}\biggr)} 
+ \f{T}{4\pi \ga} \log \biggl(\ds 1+\f{V''}{\ga \l^2 |\L|^2}\biggr)
\ee

\noi This flow equation can be rewritten in terms of the rescaled  parameters
$\ol V= \ds \f{V}{\ga \l^2  |\L|^2}$ and $\ol T= \ds \f{T}{2 \pi \ga \l^2}$. 
Putting $g= \ds \f{1}{\ga} \f{d\ga}{d\ell}$, we have:

\be\label{eq_ren12}
\f{d \ol V}{d\ell}=\ds (2-g)\ol V - \pi \f{{\ol V'}^2}{(1+{\ol V''})} +
                       \f{\ol T}{2} \log(1+{\ol V''})
\ee

\subsection{Renormalization of the surface tension $\ga$}

The contributions to the gradient term  are  obtained  from equations 
(\ref{eq_ren2}) and (\ref{eq_ren8}).

$\bullet$
Consider first the contribution due to equation (\ref{eq_ren8}).
Here, $x$ is fixed since we have imposed that 
it should be equal to $y$. We can thus expand the $V''$
terms with respect to  this variable as:

\be\label{eq_ren13}
V''\biggl(\f{\P^>(x_j)}{\l}\biggr)=V''\biggl(\f{\P^>(x)}{\l}\biggr)
+\f{1}{\l}(x_j-x) \centerdot
\na \P^>(x) V'''\biggl(\f{\P^>(x)}{\l}\biggr)
\ee

\noi  The contribution of equation (\ref{eq_ren8}) to the gradient term
is given by:

\be\label{eq_ren14}
\ba{ll}
-\ds \f{\b \ga}{2}  a^d  (-1)^{p+1} & 
\ds \biggl(\f{1}{\ga \l^2}\biggr)^{p+2}   
\sum_{n<m}  \ds \int 
\ds \prod_{j=1}^p d^d\t x_j d^dx 

\\ \\

&  (\t x_n \centerdot \na \P(x))  (\t x_m \centerdot \na \P(x)) 
\ds V'^2\biggl(\f{\P^>(x)}{\l}\biggr)
V''^{p-2}\biggl(\f{\P^>(x)}{\l}\biggr)
V'''^2\biggl(\f{\P^>(x)}{\l}\biggr)

\\ \\

& \quad  \quad \quad \quad \quad  \quad \quad
\ds \int \prod_{j=1}^p \t dk_j \t dk \
\ds \f{e^{ik \t x_1+ik_1(\t x_1-\t x_2)...+ik_p \t x_p}}
{|k|^2|k_1|^2...|k_p|^2}
\ea
\ee

\noi After integrating over $\t x_i$ for $i \neq m,n$, we are left with:

\be\label{eq_ren15}
\ba{rl}
\ds -\f{\b \ga}{2}  a^d  (-1)^{p+1} \ds \f{1}{\ga ^{p+2}} 
\ds  \sum_{n<m} & \ds  \int 
\ds d^dx d^d\t x_m d^d\t x_n  \sum_{\nu =1}^{N}
\ \t x^\nu_m  \t x^\nu_n (\na \P(x))^2_\nu

\\ \\

\ds \quad \quad \quad V'^2\biggl(\f{\P^>(x)}{\l}\biggr) & \ds
V''^{p-2}\biggl(\f{\P^>(x)}{\l}\biggr)
V'''^2\biggl(\f{\P^>(x)}{\l}\biggr)

\\ \\

& \quad \ds \int \t dk d\t k' \t dk'' \
\f{\ds e^{-ik\t x_n+ik''\t x_m+ik'(\t x_n-\t x_m)}}
{(|k|^2)^n(|k'|^2)^{m-n}(|k''|^2)^{p+1-m}}
\ea
\ee

\noi Using the fact that 
$\ds ix^{\nu}e^{ikx}=\ds \f{\pa}{\pa k^{\nu}}e^{ikx}$, 
and an integration by parts, we finally get after summing over 
$n<m$ and over $p$:

\be\label{eq_ren16}
-\f{\b \ga}{2} d\ell \  \f{4}{d} K_d (2\pi)^d  
\ds \int d^dx \
\f{\ds \ol V'^2\biggl(\ds \f{\P^<(x)}{\l}\biggr)
\ol V'''^2\biggl(\ds \f{\P^<(x)}{\l}\biggr)}
{\biggl(1+\ds \ol V''\ds \biggl(\f{\P^<(x)}{\l}\biggr)\biggr)^5}
\ds \ (\na \P^<(x))^2
\ee

\noi Since we are looking for the contribution to the gradient term, only the 
projection of the periodic function 
$\ds \f{\ol V'^2\ol V'''^2}{\biggl(1+\ol V''\biggr)^5}$
on  the zeroth harmonic counts.  
The contribution of equation 
(\ref{eq_ren8}) to the elastic constant $\ga$  is thus:

\be\label{eq_ren17}
\ds \ga d\ell \  \f{4}{d} K_d (2\pi)^d  \  \ds \int_0^{1} d\p \
\f{{\ol V}'^2(\p){\ol V}'''^2(\p)}
{\biggl(1+{\ol V}''(\p)\biggr)^5}
\ee

$\bullet$
Consider now equation (\ref{eq_ren2}). In order to obtain a term of the form 
$(\na \P^>(x))^2$, we have to expand two $V''$ terms. We proceed as follows:
if we choose to expand  
$V''(\P^<(x_n)/\l)$ and $V''(\P^<(x_m)/\l)$ with $m<n$,
we perform the expansion with 
respect to $(x_m+x_n)/2$. The $V''$ terms thus give:

\be\label{eq_ren18}
-\f{1}{4}\biggl((x_n-x_m) \centerdot 
\ds \na \P^<\biggl(\f{x_n+x_m}{2}\biggr)\biggr)^2
V'''^2\biggl(\f{x_n+x_m}{2}\biggr)
V''^{p-2}\biggl(\f{x_n+x_m}{2}\biggr)
\ee

\noi Integrating over $x_j$ for $j \neq m,n$, we get:

\be\label{eq_ren19}
\ba{ll}
-\ds \f{\b \ga}{2} & \ds \f{(-1)^p}{p} \f{T}{\ga} 
\biggl(\f{1}{\ga \l^2}\biggr)^p
\ds \sum_{m<n} \int d^dx_m d^dx_n 

\\ \\

& \ds \biggl(\na \P^<\biggl(\f{x_n+x_m}{2}\biggr)\biggr)_{\nu}^2
V'''^2\biggl(\f{x_n+x_m}{2}\biggr) 
V''^{p-2}\biggl(\f{x_n+x_m}{2}\biggr)

\\ \\

& \quad \quad \quad \quad \quad  \ds \int \t dk \t dk' \
(x_m-x_n)_{\nu}^2 \ \f{e^{i(k-k')(x_m-x_n)}}{(|k|^2)^{p-n+m}(|k'|^2)^{n-m}}
\ea
\ee

\noi Writing 
$\ds (x_m-x_n)^2_{\nu}\ \ds e^{i(k-k')(x_m-x_n)}
= \ds \f{\pa}{\pa k_{\nu}}\f{\pa}{\pa k'_{\nu}}e^{i(k-k')(x_m-x_n)}$,
and performing the rest of the calculation as described above, we find that the
contribution to the elastic term in terms of the rescaled parameters $\ol V$
and $\ol T$ is given by

\be\label{eq_ren20}
\ga d\ell \ \f{\ol T}{2d}  \int_0^{1} d\p \
\f{{\ol V}'''^2(\p)}{\biggl(1+{\ol V}''(\p)\biggr)^4}
\ee

\noi finally leading, for $d=2$, to the renormalization of $\ga$ given in the 
main text. If we had chosen another expansion scheme to obtain the 
contribution to the gradient term, for instance if we had expanded the terms
with respect to the centre of mass, we would have obtained a somewhat different 
value for $g$. This would only affect the precise value of the exponents 
$\alpha$ and $\delta$ obtained in the text, but not the qualitative features
of the solution.

\end{document}